\documentclass[preprint]{aastex}
\usepackage{emulateapj5}
\usepackage{apjfonts}
\usepackage{natbib}

\input psfig.tex

\def\aa{{A\&A}}
\def\aas{{ A\&AS}}
\def\aj{{AJ}}
\def\al{$\alpha$}
\def\bet{$\beta$}

\def\annrev{{ARA\&A}}
\def\apj{{ApJ}}
\def\apjs{{ApJS}}

\def\deg{$^{\circ}$}

\def\kms{km s$^{-1}$}
\def\lamb{$\lambda$}
\def\lax{{$\mathrel{\hbox{\rlap{\hbox{\lower4pt\hbox{$\sim$}}}\hbox{$<$}}}$}}
\def\gax{{$\mathrel{\hbox{\rlap{\hbox{\lower4pt\hbox{$\sim$}}}\hbox{$>$}}}$}}
\def\simlt{\lower.5ex\hbox{$\; \buildrel < \over \sim \;$}}
\def\simgt{\lower.5ex\hbox{$\; \buildrel > \over \sim \;$}}

\def\msigma{{$M_{\bullet}-\sigma_*$}}
\def\micron{{$\mu$m}}
\def\mnras{{MNRAS}}

\def\percm2{cm$^{-2}$}

\def\solum{$L_\odot$}
\def\solmass{$M_\odot$}

\def\hi{\ion{H}{1}}

\def\oiii{[\ion{O}{3}]}

\def\vm{${\upsilon_m}$}

\def\sig{$\sigma_0$}
\def\sigstar{$\sigma_*$}

\slugcomment{To appear in {\it The Astrophysical Journal}.}
\lefthead{Ho}
\righthead{High-$z$ Quasar Host Galaxies}

\begin{document}

\title{The CO Tully-Fisher Relation and Implications for the Host Galaxies 
of High-Redshift Quasars}

\author{Luis C. Ho}

\affil{The Observatories of the Carnegie Institution of Washington, 813 Santa 
Barbara St., Pasadena, CA 91101}

\begin{abstract}
The integrated line width derived from CO spectroscopy provides a powerful 
tool to study the internal kinematics of extragalactic objects, including 
quasars at high redshift, provided that the observed line width can be 
properly translated to more conventionally used kinematical parameters of 
galaxies.  We show, through construction of a $K_s$-band CO Tully-Fisher 
relation for nearby galaxies spanning a wide range in infrared luminosity,
that the CO line width measured at 20\% of the peak intensity, when 
corrected for inclination and other effects, successfully recovers the maximum 
rotation velocity of the disk.  The line width at 50\% of the peak intensity 
performs much more poorly, in large part because CO lines have a 
wide range of profiles, which are shown to vary systematically with infrared 
luminosity.  We present a practical prescription for converting observed CO 
line widths into the stellar velocity dispersion of the bulge (\sigstar), and 
then apply it to a sample of low-redshift ($z$ \lax\ 0.2) and high-redshift  
(1.4 \lax\ $z$ \lax\ 6.4) quasars to study their host galaxies.  Nearby quasars 
roughly fall on the correlation between black hole mass and bulge stellar 
velocity dispersion established for inactive galaxies, but the host galaxies 
of the high-$z$ quasars systematically deviate from the local 
$M_{\bullet}-\sigma_*$ relation.  At a given $\sigma_*$, high-$z$ quasars have 
black hole masses larger by a factor of $\sim$4 relative to local galaxies, 
suggesting that early in the life-cycle of galaxies the development of the 
bulge lags behind the growth of the central black hole.  An alternative 
explanation for these observations, which currently cannot be ruled out 
rigorously, is that high-redshift quasars are preferentially viewed at 
face-on orientations.
\end{abstract}


\section{Introduction}

The discovery of strong empirical relations between black hole (BH) mass and 
the properties of nearby galaxies, both inactive (Kormendy \& Richstone 1995; 
Magorrian et al. 1998; Gebhardt et al. 2000a; Ferrarese \& Merritt 2000) and 
active (Gebhardt et al. 2000b; Ferrarese et al. 2001; Barth et al. 2005; 
Greene \& Ho 2006), has triggered an upsurge of interest in the coevolution of 
BHs and galaxies (e.g., Ho 2004).   An important unanswered question is when 
the local scaling relations were established.  While there is no shortage of 
opinions from theoretical studies (e.g., Croton 2006; Fontanot et al. 2006; 
Hopkins et al. 2007), to date solid empirical evidence has been relatively 
meager, and somewhat conflicting.  From stellar velocity dispersion 
measurements of moderate-luminosity active galactic nuclei (AGNs), Treu et al. 
(2004, 2007) and Woo et al. (2006) find that the \msigma\ relation at $z=0.36$ 
already begins to show departure from the local relation.  On the other hand, 
studies of larger and more luminous AGN samples, but using less direct 
surrogates of $\sigma_*$ based on the widths of optical nebular emission lines,
find little or no evolution of the \msigma\ relation out to $z \approx 1.2$ 
(Salviander et al. 2007) and $z \approx 3$ (Shields et al. 2003).  At even 
higher redshifts, Walter et al. (2004) resolved the molecular gas in the 
$z = 6.42$ quasar SDSS~J114816.64+525150.3 into an apparent nuclear disk, from 
which they were able to derive a dynamical mass that, taken at face value, 
indicates that the host galaxy has an embryonic bulge compared to the estimated 
mass of the BH.  Wei\ss\ et al. (2007) arrived at a very similar conclusion 
from their detailed study of the $z = 3.91$ lensed quasar APM~08279+5255.  
While the interpretation of these results is not unambiguous, it qualitatively 
agrees with the photometric and lensing studies of high-$z$ quasars by Peng et 
al. (2006a, 2006b), which show that at 1\lax\ $z$ \lax\ 4.5 the ratio of BH 
mass to host galaxy mass is a factor of several higher than the value in the 
local Universe. 

The masses of high-$z$ quasar host galaxies are challenging to estimate 
accurately, and likely will remain so for some time.  To make progress one 
should explore multiple avenues.  One promising technique is to exploit the
kinematic information encoded in molecular line observations.  While studying
molecular gas in high-$z$ sources is still far from routine, an 
increasing number of detections have been reported (Solomon \& Vanden~Bout 
2005, and references therein), and the upcoming generation of new 
millimeter-wave facilities hold great promise for rapid progress in this area.
Shields et al. (2006) recently attempted to constrain the host galaxy masses of 
nine high-$z$ quasars with available measurements of the CO line width and 
BH mass.  In order to place the high-$z$ objects on the \msigma\ relation, they 
calibrated the width of the CO line profile at 50\% of the peak intensity 
(hereafter $W_{50}$) against a small sample of low-$z$ quasars for which 
they could obtain indirect estimates of the bulge stellar velocity dispersion, 
concluding in the end that $\sigma_*$ could be estimated simply from 
$W_{50}$/2.35.  Since the currently available high-$z$ quasars have 
conspicuously narrow CO lines (Greve et al. 2005; Carilli \& Wang 2006), 
Shields et al.  (2006) suggest that these objects deviate strongly from the 
local \msigma\ relation in having a much larger BH mass for a given 
$\sigma_*$.  The implication is that the growth of the BH predates, and is 
largely decoupled from, the formation of the galaxy bulge.

Although the interpretation of Shields et al. (2006) resonates well with the 
findings of Peng et al. (2006a, 2006b), it is not without complications.   With 
a few exceptions (e.g., SDSS~J114816.64+525150.3; Walter et al. 2004), most 
of the other quasars are not spatially resolved, and thus we have no knowledge 
of the spatial distribution of the molecular gas.  Apart from signifying a 
shallow gravitational potential, a narrow CO line can arise from at least two 
other effects.  First, the CO may be sufficiently concentrated toward the 
center of the host that it only samples the innermost part of the galaxy's 
rotation curve, thereby underestimating the maximum rotation velocity.  This is 
well-known from observations (e.g., Downes \& Solomon 1998) as well as 
numerical simulations (e.g., Barnes \& Hernquist 1996) of merging galaxies, 
wherein the bulk of the molecular gas, as a result of tidal torques, sinks to 
the central few hundred parsecs and forms a compact nuclear disk or ring.  As 
the hosts of high-$z$ quasars likely involve mergers, this should be a source 
of concern.  Second, as argued by Carilli \& Wang (2006), the distribution of 
inclination angles for the CO-emitting disk of high-$z$ quasars, by virtue of 
their luminosity bias and selection method, may preferentially favor more 
face-on orientations.  If so, then the inclination correction to the CO line 
widths can be arbitrarily large.  For the high-$z$ sample to conform to the 
local \msigma\ relation, the average inclination angle needs to be $\langle i 
\rangle \approx 13-15$\deg\ (Carilli \& Wang 2006; Wu 2007).  Lastly, 
we note that the interpretation of these and future observations depends quite 
critically on how one actually translates the CO line width into a more 
familiar dynamical variable.  Shields et al. (2006) chose to calibrate 
$W_{50}$ with three indirect estimators of $\sigma_*$ (virial BH masses and
\oiii\ \lamb 5007 line widths for seven objects, and host galaxy luminosities 
for two objects).  By contrast, Wu (2007) examined directly the correlation 
between $W_{50}$ and $\sigma_*$ for a significantly larger sample of 33 Seyfert 
galaxies, concluding that the Shields et al. calibration is significantly 
nonlinear.

This paper presents a new method of estimating $\sigma_*$ based on more 
complete information on the CO line profile.  We begin by asking three basic 
questions that we believe have not been adequately addressed.  (1) How well 
does the CO line width trace the kinematics of galaxies and precisely what part 
of the galaxy does it probe?  (2) What part of the CO profile, which, unlike 
that of \hi\, is often not sharp and double-horned, should one measure?  
And (3) how do we relate the CO line width to the kinematics of the 
bulge?   We investigate the first two issues by reexamining the CO Tully-Fisher 
relation, with particular emphasis on infrared-luminous galaxies, perhaps 
the best local analogs of what high-$z$ quasar host galaxies might look like.
We find that galaxies over a wide range of infrared (IR) luminosities fall on 
the CO Tully-Fisher relation provided that one uses $W_{20}$, the width of the 
CO line at 20\% of the peak intensity, rather than $W_{50}$.  Moreover, we show 
that the shape of the line profile, as parameterized by the ratio 
$W_{20}$/$W_{50}$, varies significantly and systematically with IR luminosity.
This thus provides a simple prescription for converting $W_{50}$ to the 
preferred parameter $W_{20}$ if the IR luminosity is known.  Having shown that
the CO line width can effectively probe the maximum rotation velocity (\vm) of 
the galaxy, we then use the new calibration of the \vm--\sigstar\ relation 
by Ho (2007) and discuss the implications for high-$z$ quasar host galaxies.

\vskip 1.5cm
\section{The CO Tully-Fisher Relation for \\ Infrared-luminous Galaxies}

The CO Tully-Fisher relation is analogous to the more familiar relation 
commonly used with \hi\ (Tully \& Fisher 1977).  Originally proposed by Dickey 
\& Kas\'es (1992) and Sofue (1992), it has been applied to normal galaxies both 
nearby (Sch\"oniger \& Sofue 1994, 1997; Lavezzi \& Dickey 1998) and at 
moderate distances up to $z \approx 0.1$ (Sofue et al. 1996; Tutui et al. 
2001).  As discussed by these authors, CO offers a number of potential 
advantages compared to \hi, chief among them being the relative insensitivity 
of the CO distribution to environment perturbations and the technical 
feasibility of observing to greater distances.  

One of the practical difficulties of using the CO line stems from the wide 
variety of line profiles actually observed.  Unlike \hi, which generally 
exhibits a sharp-edged, double-horned profile characteristic of an optically 
thin rotating disk, CO line profiles are much more irregular, with shapes 
ranging from classic double peaks, to more rounded single peaks, to near 
Gaussians (e.g., Sanders et al. 1991; Dickey \& Kas\'es 1992).  As discussed in 
Lavezzi \& Dickey (1997), many factors contribute to this diversity in CO line 
profiles, two physically interesting ones being the radial distribution of 
the gas and its opacity.  In spiral galaxies the molecular gas is generally 
more centrally concentrated than the atomic hydrogen, and the $^{12}$CO(1--0) 
transition tends to be optically thick.  Nevertheless, in practice the 
full-width near zero intensity (or approximately $W_{20}$) of the CO line quite 
faithfully follows that of \hi\ (Dickey \& Kas\'es 1992; Sch\"oniger \& Sofue 
1994).
The molecular gas distribution, although more concentrated than the \hi, 
evidently extends sufficiently past the peak of the rotation curve to imprint 
the maximum rotation velocity on the CO line profile.  To achieve the best 
possible match between CO and \hi\ velocities, one should apply corrections for 
interstellar turbulence (Lavezzi \& Dickey 1997), as well as accounting for the 
relative differences between the radial distribution of the two tracers and 
luminosity-dependent differences in the gradient of the rotation curve (Tutui 
\& Sofue 1999).

Do quasar host galaxies also obey the CO Tully-Fisher relation?  Insofar 
as galaxy interactions or mergers are often implicated as the triggering 
mechanism of quasars (e.g., Hopkins et al. 2006), which are accompanied or 
preceded by starburst activity  (e.g., Ho 2005), this issue can be effectively
addressed by looking into the properties of nearby IR-luminous galaxies 
(Sanders \& Mirabel 1996).  We have assembled a large sample of galaxies 
spanning a wide range in IR luminosities\footnote{Throughout this paper 
$L_{\rm IR}$ refers to the total IR (8--1000 \micron) luminosity, derived using 
12, 25, 60, and 100 \micron\ flux densities from the {\it Infrared Astronomical 
Satellite (IRAS)} following the prescription of Sanders \& Mirabel (1996). The
{\it IRAS}\ data are taken from the NASA/IPAC Extragalactic Database (NED; 
{\tt http://nedwww.ipac.caltech.edu}).} that contain measurements of both 
$W_{20}$ and $W_{50}$ for the $^{12}$CO(1--0) line integrated over at least a 
central diameter of $\sim$10 kpc, but preferably covering the entire galaxy.  
This last criterion is very important because most nearby galaxies are larger 
than the beam sizes of millimeter-wave telescopes, whereas we wish to 
simulate, as closely as possible, the global measurements typically achieved 
with single-dish \hi\ telescopes in order to capture the full extent of the 
rotation curve.  The rotation curves of spiral galaxies usually reach their 
maximum velocity at $r \approx 2-7$ kpc, beyond which they remain flat (Sofue 
\& Rubin 2001).  To achieve this, most of the objects are relatively distant 
(60\% are beyond 70 Mpc); more nearby systems have been mapped, and the 
published line widths represent the synthesized profile over 

\begin{figure*}[t]
\centerline{\psfig{file=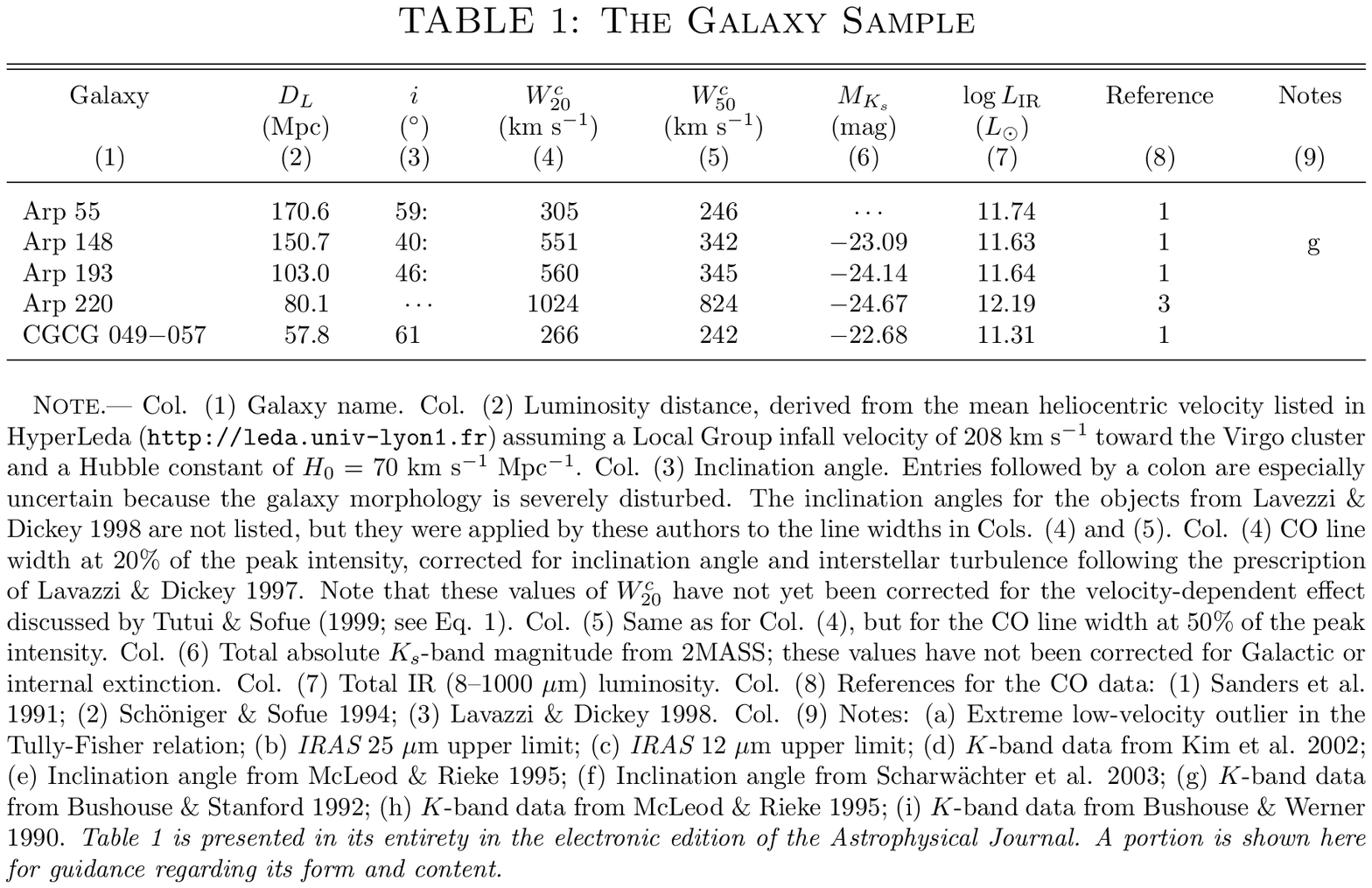,width=18.5cm,angle=0}}
\end{figure*}

\noindent
the mapped 
positions.  The final sample of 151 galaxies, assembled in Table~1, covers 
$L_{\rm IR} \approx 10^8$ to $4\times 10^{12}$ \solum, ranging from nearby 
normal galaxies to those of the ``luminous IR'' and ``ultraluminous IR'' 
variety.  The data were taken from three compilations: (1) luminous {\it IRAS}\ 
galaxies from Sanders et al. (1991); (2) nearby galaxies from Sch\"oniger \& 
Sofue (1994), which in addition contains the two powerful AGNs Mrk 231 and 
I Zw 1; and (3) the distant cluster members and ultraluminous IR galaxies 
from Lavezzi \& Dickey (1998).  

Inclination angles were estimated from axial ratios obtained mostly from 
de~Vaucouleurs et al. (1991), and otherwise from Paturel et al. (2000), as 
described in Ho (2007), except that the intrinsic thickness of the disk was 
fixed at a value of 0.2 instead of allowing it to vary with Hubble type.  The 
morphologies of many of the IR-luminous galaxies from the sample of Sanders et 
al. (1991) are particularly difficult to define, as a number of them are 
merger remnants or otherwise appear tidally disturbed.  In Table~1 we have 
flagged the most suspicious values with a colon.  To avoid excessively large 
and uncertain inclination corrections, we excluded objects with inclination 
angles less than 30\deg, except in the case of the Sanders et al. (1991) 
sample on account of the inherent difficulty of obtaining accurate inclinations 
for many of these sources.  Following Lavezzi \& Dickey (1997), we 
apply a correction for turbulence broadening to the CO line widths, adopting 
their formalism and their recommended value of 38 \kms\ and 19 \kms\ for the 
correction to $W_{20}$ and $W_{50}$, respectively.  [We did not apply 
corrections for inclination or turbulence broadening to the Lavezzi \& Dickey 
(1998) sample because these authors had already done so.]    Lastly, we also
apply the velocity-dependent correction recommended by Tutui \& Sofue (1999) 
to make $W_{20}$ fully compatible with the \hi-based velocities.  To account 
for the different radial distribution of CO versus \hi\ and for variations in 
the gradient of the rotation curve with luminosity (and hence rotation 
velocity), these authors suggest the following calibration:

\begin{equation}
W_{20}^c = 0.76\ W_{20} + 83.8.
\end{equation}

\vskip 0.2cm
\noindent
We confirm that applying this additional correction does result in marked 
improvement in the final CO Tully-Fisher relation.

Figure~1 shows the $K_s$-band Tully-Fisher relation for the sample, plotting 
separately the maximum rotation velocity as estimated from the 
corrected CO line widths measured at 50\% ($W_{50}^c$) and 20\% 
($W_{20}^c$) of the line peak, and dividing the sample into normal and 
IR-luminous galaxies.  With just a few exceptions, the $K_s$-band (2.16 
\micron) magnitudes come from homogeneous photometry from the Extended Source 
Catalog of the Two-Micron All-Sky Survey (2MASS; Skrutskie et al. 2006).  
Overplotted on the figure is the \hi\ Tully-Fisher relation derived by 
Verheijen (2001) for galaxies in the Ursa Major cluster in the $K^\prime$ 
band, which is quite similar to the 2MASS $K_s$ band (Bessel 2005).  Looking 
first at the normal galaxies (top two panels), it is clear that $W_{20}^c$ 
fares better than $W_{50}^c$ in matching Verheijen's relation.  Not only does 
the zeropoint agree better, the scatter is also reduced, from a 
root-mean-square (rms) of 1.0 mag to 0.77 mag.  A large part of the
improvement stems from the fact that the Tutui-Sofue correction has been 
applied to $W_{20}$ but not to $W_{50}$ (because it is not currently known for 
the latter); without this  correction, the scatter reduces only to an rms of 
0.93 mag.  The observed scatter for our sample is larger than 
Verheijen's, but this is hardly surprising considering the larger distance 
errors and greater heterogeneity of our sample.  The bulk of the objects lie 
within the boundaries that enclose twice the rms 

\vskip 0.3cm
\begin{figure*}[t]
\centerline{\psfig{file=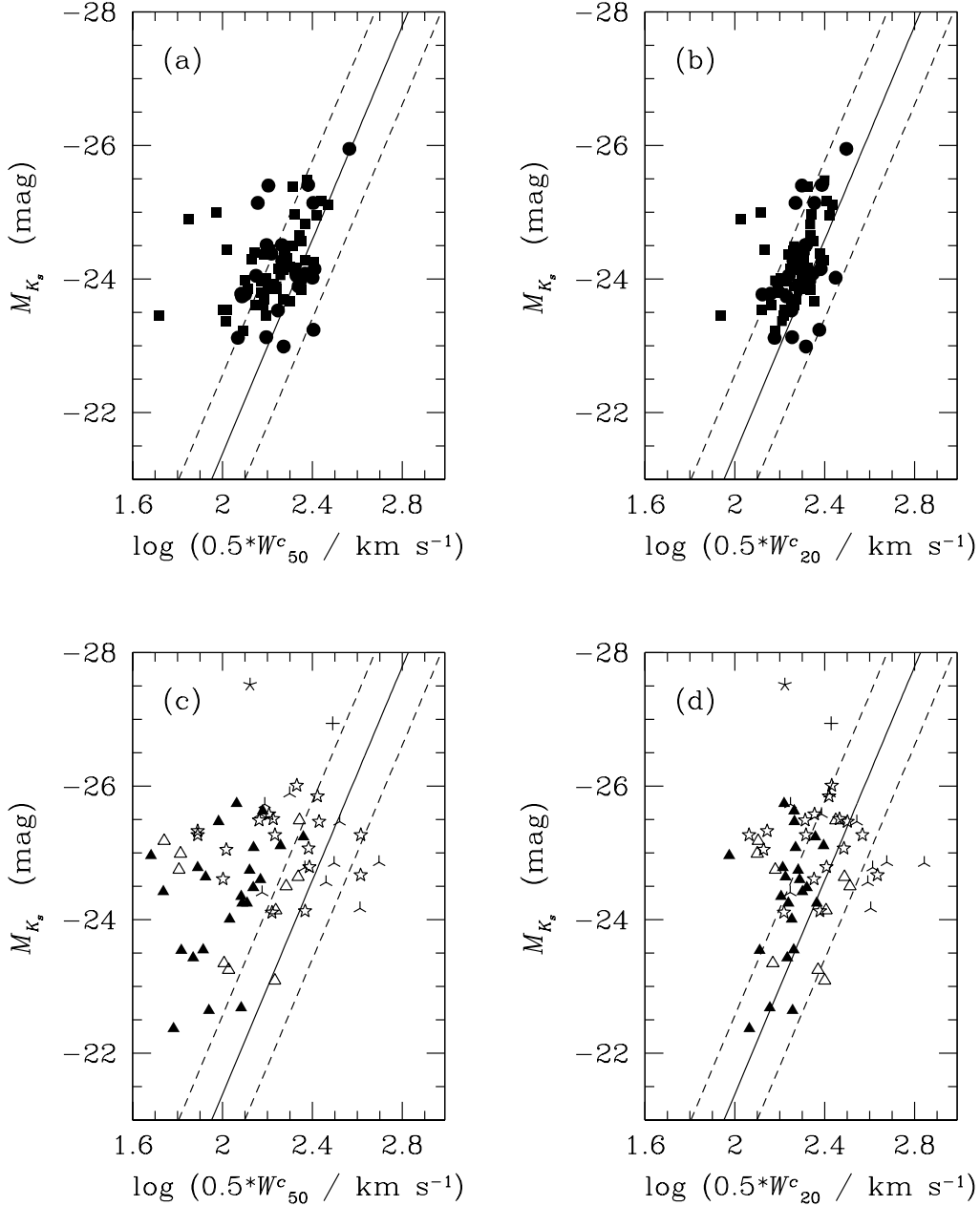,width=14.5cm,angle=0}}
\figcaption[f1.eps]{
The CO Tully-Fisher relation in the $K_s$ band for ({\it top}) normal galaxies
and ({\it bottom}) IR-luminous and IR-ultraluminous galaxies.  The maximum
rotation velocity is estimated using ({\it left}) $\onehalf W_{50}^c$ and
({\it right}) $\onehalf W_{20}^c$.  The values of $W_{50}^c$ and $W_{20}^c$
have been corrected for inclination and turbulence, and $W_{20}^c$ has been
corrected in addition for the Tutui-Sofue (1999) effect (see Eq.~1).  In
the top panels the circles are the nearby galaxies from Sch\"oniger \& Sofue
(1994), and the squares are the cluster galaxies from Lavezzi \& Dickey
(1998); only objects with $i \geq 30$\deg\ are included.  In the bottom
panels, the triangles come from the sample of Sanders et al. (1991), where the
solid symbols represent objects without obvious optical morphological
peculiarities, open symbols represent those with clearly disturbed
morphologies or tidal features, and skeletal symbols denote those with
inclination angles formally less than 30\deg.  The open stars mark the 21
ultraluminous IR galaxies from the study of Solomon et al. (1997) for which
Lavezzi \& Dickey (1998) obtained follow-up optical imaging to determine
inclination angles.  Mrk 231 and I Zw 1 from Sch\"oniger \& Sofue (1994) are
plotted as an asterisk and a plus symbol, respectively. The {\it solid line}\
shows the $K^\prime$-band Tully-Fisher relation from Verheijen (2001) based on
\hi\ rotation curves of spiral galaxies in the Ursa Major cluster; the 
{\it dashed lines}\ lines mark the region that has twice the rms scatter of 
the Ursa Major objects.
\label{f1}}
\end{figure*}
\vskip 0.3cm

\noindent
scatter of Verheijen's
``\hi'' sample (0.59 mag).  There appears to be a small zeropoint offset of
$\sim 0.3$ mag (or $\sim 9$\% increase in velocity), but given the limited
size and character of our sample, we regard this finding as tentative.
Overall, our analysis reaffirms the results of previous attempts to compare
the CO and \hi\ Tully-Fisher relations at optical wavelengths, now extended
for the first time to the near-IR.

That $W_{20}^c$ is far superior than $W_{50}^c$ in recovering the maximum
rotation velocity is most dramatically illustrated for the sample of
IR-luminous galaxies shown on the bottom two panels of Figure~1.  Employing
$W_{50}^c$ places the bulk of the objects significantly offset (by $\sim 2$
mag), with enormous scatter (rms = 1.9 mag), to the low-velocity side of the
reference Tully-Fisher locus. The use of $W_{20}^c$, on the other hand, shifts 
most of the objects back to 
the fiducial ridgeline and markedly tightens the distribution (rms = 1.5 mag).
Again, the Tutui-Sofue correction to $W_{20}$ has a significant impact; had it 
not been applied, the rms scatter is reduced only to 1.7 mag.  To be sure, the 
scatter is larger than in the case of the normal galaxies, but this is 
entirely to be expected given how poorly we know the inclination angles, not 
to mention of the strong tidal perturbations witnessed in the stars, which 
presumably must also be imprinted to some degree on the gas.  There are no 
clear-cut differences when the sample is separated either by the degree of 
morphological peculiarity as judged by our qualitative visual examination of 
optical images in NED ({\it open}\ versus {\it solid triangles}) or by the more 
detailed ``interaction types'' given in Sanders et al. (1991), nor do the 
highest luminosity sources ({\it open stars}) stand out in any conspicuous 
manner.  Despite the appreciable star formation rates in many of these 
systems, most of which qualify as hefty starbursts, we see no luminosity 
offset relative to the fiducial ridgeline.  The modest amount of brightening 
predicted in the $K$ band for starbursts ($\sim 0.5$ mag; Mouhcine \& 
Lan\c{c}on 2003) may be imperceptible in the midst of the large scatter in the 
figure.

As an aside, we note that our sample of normal galaxies (Fig.~1{\it b}) 
contains 3--4 outliers with exceptionally low rotation velocities, reminiscent 
of the population of kinematically anomalous galaxies identified by Ho (2007) 
on the basis of \hi\ measurements.  Ho posits that these objects possess
substantial amounts of dynamically unrelaxed atomic hydrogen captured 
either from minor mergers or possibly accreted directly from primordial
clouds.  However, in view of the substantial quantity of molecular gas 
involved in the current sample, a primordial origin seems far-fetched.  
Instead, we may appeal to the other arguments presented by Ho (2007), namely 
that the gas and stars might be strongly misaligned, or else the gas has been 
severely perturbed by tidal interactions.  Since all of the outliers belong to 
the cluster sample of Lavezzi \& Dickey (1998), this is perhaps 
not implausible.  The IR-luminous subsample (Fig.~1{\it d}) also contains a 
number of low-velocity outliers, but these are balanced by an almost equal 
number of high-velocity outliers, a situation that can be explained by the 
gas and stars not being coplanar (Ho 2007).  Given the great difficulty in 
estimating inclination angles for this subsample, however, we should not 
attach too much weight to the observed scatter.

In the context of this study, the critical point of the above analysis is that
the CO line width, as measured by $W_{20}^c$, provides a robust measurement of
the maximum rotation velocity of the large-scale disk of the galaxy, on nearly 
equal footing with velocities more conventionally derived from integrated \hi\ 
line widths or extended \hi\ or H\al\ rotation curves.  This holds not only for
normal galaxies, but also for IR-luminous systems, whose evolutionary status
may be hoped to bear some resemblance to that of quasar host galaxies.

\vskip 0.5cm
\section{Estimating \sigstar\ from CO Line Widths}

We have established that the width of the CO line measured at 20\% of the peak 
intensity, $W_{20}^c$, accurately recovers the maximum rotation velocity 
of the disk, \vm.  On the other hand, the kinematic quantity most closely 
linked to BH mass is the velocity dispersion of the bulge, \sigstar, not \vm.  
A number of studies (e.g., Ferrarese 2002; Pizzella et al. 2005) have 
suggested that galaxies over a wide range of morphological types obey a tight, 
nearly linear correlation between \vm\ and \sigstar, and by virtue of 
the \msigma\ relation, that BH mass may be equally, or perhaps even more 
fundamentally, coupled to \vm\ than to \sigstar.  The recent analysis of Ho 
(2007; see also Courteau et al. 2007), however, shows that in fact the 
\vm--\sigstar\ relation exhibits substantial intrinsic scatter and that its 
zeropoint varies systematically with galaxy morphology, bulge-to-disk
ratio, and light concentration.  Despite these complications, \vm\ {\it is}\ 
correlated with \sigstar, both as observed, not only for inactive but also 
active galaxies (Ho et al. 2007), and as anticipated from basic dynamical 
considerations (see discussion in Ho 2007), and so we can take 
advantage of this fact to translate the CO line width into \sigstar.  
Ho (2007) presented fits to the \vm--\sig\ relation using the largest and 
most comprehensive calibration sample to date.  Since the zeropoint depends on 
morphological type, and the scatter is somewhat reduced for later-type 
galaxies, we use here the fit for the 550 ``kinematically normal'' spiral 
galaxies in Ho's sample (Eq.~5 in Ho 2007), to yield

\begin{equation}
\log \sigma_* = (1.26\pm0.046) \log \upsilon_m - (0.78\pm0.11),
\end{equation}

\vskip 0.2cm
\noindent
where in the present context $\upsilon_m \approx \onehalf W_{20}^c$. 

Despite the advantages of $W_{20}^c$ over $W_{50}^c$, two practical 
difficulties often preclude using the former over the latter.  The line 
width near the base of the profile is maximally sensitive to the noise in 
the spectrum and to errors in baseline determination, and hence $W_{20}$ 
is inherently more uncertain than $W_{50}$.  Moreover, many studies, 
presumably for this very reason, do not even publish $W_{20}$, but $W_{50}$ is 
often given.  Now, $W_{20}$ can be estimated from $W_{50}$ in two simple
cases: for a sharp-edged, double-horned profile $W_{20}/W_{50} \approx 1$, and 
for a Gaussian $W_{20}/W_{50} \approx 1.5$.  Many CO profiles, however, do not 
conform to these two simple shapes, and indeed real line profiles can even 
have wings more extended than a Gaussian, and so it would be highly desirable 
to devise an empirical guideline to estimate $W_{20}$ from $W_{50}$.  Note that 
the differences between these two examples have dramatic consequences for 
the inferred BH masses.  A 50\% increase in the CO line width, for example, 
when folded through Equation~2 and the \msigma\ relation of Tremaine et al. 
(2002; $M_\bullet \propto \sigma_*^{4.02}$), results in a factor of $\sim 8$ 
increase in the predicted BH mass.

Recall from our discussion of Figure~1 that the impact of substituting 
$W_{50}^c$ with $W_{20}^c$ had a much more dramatic effect in recovering the 
\hi\ Tully-Fisher relation in the case of the IR-luminous galaxies than for 
normal galaxies.  The implication is that the line shape must vary 
systematically with IR luminosity.  We illustrate this effect in Figure~2, 
where we plot $W_{20}^c$/$W_{50}^c$ as a function of the total IR luminosity.
At $L_{\rm IR}$ \lax\ $10^{10}$ \solum, $W_{20}^c$/$W_{50}^c \approx 1$, 
but $W_{20}^c$/$W_{50}^c$ progressively increases toward higher luminosities, 
such that by $L_{\rm IR} \approx 10^{12}$ \solum\ it attains a median value 
close to 1.5, the expected value for a Gaussian adopted by Shields et al. 
(2006).  Despite the large scatter and the fact that 
$W_{20}^c$/$W_{50}^c$ is bounded on one side, a generalized Spearman rank-order 
correlation test formally yields a correlation coefficient of $\rho = 0.48$, 
with a probability of $< 10^{-4}$ for rejecting the null hypothesis of no 
correlation.  A linear regression fit with $\log L_{\rm IR}$ as the 
independent variable gives

\begin{equation}
W_{20}^c/W_{50}^c = (0.21\pm0.041) \log L_{\rm IR} - (1.05\pm0.44).
\end{equation}

\vskip 0.2cm
\noindent
Note that at any given luminosity there are objects spanning the full range 
of $W_{20}^c$/$W_{50}^c$, from $\sim 1$ up to a maximum value that increases 
roughly monotonically with luminosity.  It is the 
%
%
upper envelope of the 
$W_{20}^c$/$W_{50}^c$ distribution, and consequently the 

\vskip 0.3cm
\psfig{file=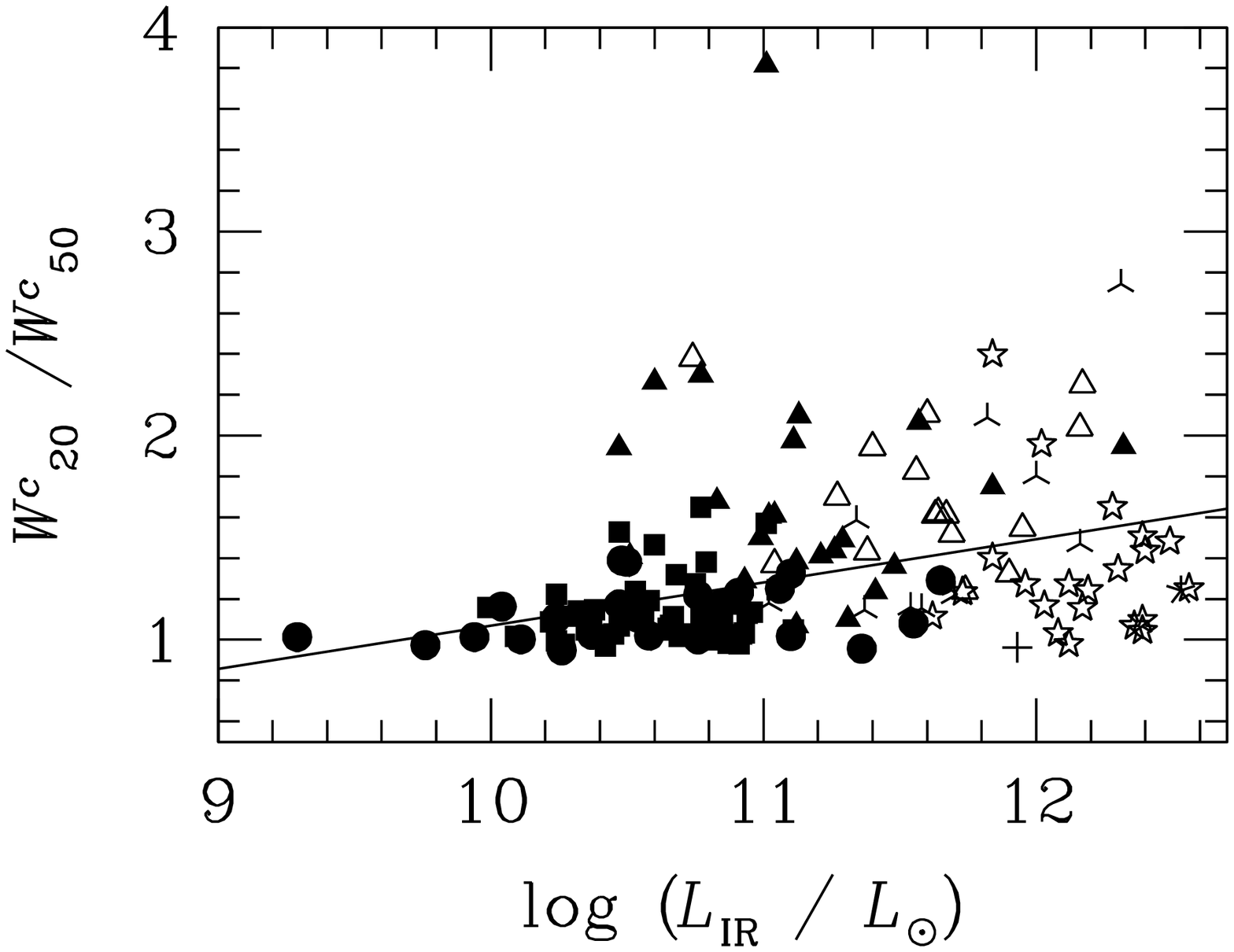,width=8.5cm,angle=0}
\figcaption[f2.eps]{
Variation of the line shape, as parameterized by the ratio
$W_{20}^c$/$W_{50}^c$, as a function of total IR luminosity.  The symbols have
the same meaning as in Figure~1.  The solid line shows the linear regression
fit between $W_{20}^c$/$W_{50}^c$ and $\log L_{\rm IR}$ (Eq.~3).
\label{f2}}
\vskip 0.3cm

\noindent
dispersion about the 
mean, that increases with luminosity; at the highest luminosities 
$W_{20}^c$/$W_{50}^c$  can be as large as 2--2.5.  Since physically $W_{20}^c$ 
should be larger than $W_{50}^c$, even if measurement errors may result 
in $W_{20}^c$/$W_{50}^c$ \lax\ 1, Equation~3 strictly speaking should only 
be applied when $W_{20}^c$/$W_{50}^c > 1$.  Also, it is worth stressing that 
the correction recommended in Equation~3 is statistical in nature; given the 
large scatter about the mean relation, especially at high luminosities, 
applying the correction to any individual source is subject to a large 
uncertainty.

Interestingly, Lavezzi \& Dickey (1997) searched for possible correlations of 
CO line shape with galaxy properties but found none that was especially 
compelling.  It could be that the effect described here was masked by their 
smaller sample and more limited dynamic range.  Multiple factors influence the 
observed CO line profile (Dickey \& Kas\'es 1992; Lavezzi \& Dickey 1997), but 
chief among them are the optical depth and the radial distribution of the 
gas.  The optical depth of the $^{12}$CO(1--0) line in starbursts, however, if 
anything, appears to be {\it lower}\ than in normal galaxies (e.g., Aalto et 
al. 1995; Glenn \& Hunter 2001), in which case Figure~2 actually exhibits the 
{\it opposite}\ trend predicted if opacity were the principal cause for the 
line shape variation.  On the other hand, it is entirely natural to expect the 
molecular gas to be more centrally concentrated in starburst galaxies.  The 
most powerful starbursts seem to arise from central stockpiles of molecular gas 
that have been dissipated to the nuclear region, often with the aid of tidal 
interactions (e.g., Barnes \& Hernquist 1996; Sanders \& Mirabel 1996; 
Downes \& Solomon 1998).

In summary, we recommend the following recipe for converting an observed CO 
line width into \sigstar.  This recipe is empirically motivated by the success 
with which we were able to recover the rotational velocity of IR-luminous 
galaxies, likely good analogs of quasars, as described in Section 2.  If the 
line width at 20\% of the peak can be reliably measured, then $W_{20}$ should 
be used, after correcting it for turbulence as described in Lavezzi \& Dickey 
(1997), inclination, and the Tutui-Sofue (1999) effect (Eq.~1).  If only the 
width at 50\% of the peak is known or can be measured with confidence, then 
after correcting for turbulence and inclination, Equation~3 should be used in 
concert with an estimate of $L_{\rm IR}$ to convert $W_{50}^c$ to $W_{20}^c$, 
followed by an adjustment for the Tutui-Sofue (1999) effect.  If the 
inclination angle of the source is unknown, as is almost always the case in 
high-$z$ objects, but can be assumed to be random, then 
$\langle i\rangle =45$\deg.  With $W_{20}^c$ at hand, \sigstar\ follows from 
the \vm-\sigstar\ relation (Eq.~2).

\vskip 0.8cm
\section{The Host Galaxies of High-$z$ Quasars}

We apply the recipe outlined above to reassess the \msigma\ relation of
high-$z$ quasars.  For comparison, we also evaluate all the low-$z$ quasars 
with available CO line widths.  The data, compiled in Table~2, are similar to 
those presented in Shields et al. (2006), but with several important 
modifications.  First, we expanded the sample of low-$z$ Palomar-Green (PG) 
quasars from 12 to 17.  The five additional sources have uniformly measured 
broad H\bet\ line widths from Boroson \& Green (1992) and optical 
spectrophotometry from Neugebauer et al. (1987), and so BH masses can be 
derived for them using the conventional virial method.  The line widths 
from the original sample of Evans et al. (2001) have now been updated by 
Evans et al. (2006).  For ease of comparison, we adopt the particular virial 
relation used by Shields et al., as well as their choice of cosmological 
parameters to derive distances.  One key difference lies in the treatment of 
the CO line widths.  The line widths for the nine objects published in 
Scoville et al. (2003) pertain to the full-width near zero intensity 
($\sim W_{20}$), not the full-width at half maximum ($W_{50}$).  Shields et 
al. obtained the $W_{50}$ values of these sources by assuming that $W_{50} = 
\twothirds W_{20}$.  Since the formalism developed in this paper explicitly 
accounts for line shape variations, Table~2 carefully documents the original 
data.  We also found it necessary to adjust the line width data for a few of 
the high-$z$ quasars. 

Figure~3 summarizes the CO-derived \msigma\ relation for quasars, where we 
have applied a correction for an average inclination angle of 45\deg.  The 
low-$z$ PG quasars roughly center around the local \msigma\ relation of 
inactive galaxies (Tremaine et al. 2002), but the scatter is very large.  
Taken at face value, the PG quasars actually lie slightly {\it below}\ the 
local relation, reminiscent of a similar tendency reported in other broad-line 
AGNs (Greene \& Ho 2006).  We refrain from quantifying the exact magnitude of 
the offset, as it depends on the choice of the virial relation used to obtain 
the BH mass, and because the current sample is both small and compromised by 
the lack of inclination angle information.  Two of the objects have 
exceptionally narrow CO lines (PG~0838+770, $W_{50} = 60$ \kms; PG~1415+451, 
$W_{50} = 90$ \kms), but they should not be viewed as too alarming, since 
similar cases are seen among inactive normal and IR-luminous galaxies.  Both 
of these quasars possess luminous hosts ($M_K \approx -25.8$ and $-24.7$ mag, 
respectively, assuming $H-K \approx 0.3$ mag; McLeod \& Rieke 1994), which 
would place them among the kinematically anomalous members in the CO 
Tully-Fisher relation (Fig.~1{\it b}).

More interesting is the situation for the high-$z$ quasars.  Consistent with 
the analysis of Shields et al. (2006), high-$z$ quasars seem to show a clear 
statistical offset above the local \msigma\ relation.  Our treatment of the
CO line widths, however, has reduced considerably the magnitude of the offset, 
from an average value of $\Delta \log M_{\rm BH} = 1.8$ dex as found by 
Shields et al. to $\Delta \log M_{\rm BH} \approx 0.6$ dex.  This brings these 
results into better agreement with the findings of Peng et al. (2006b), whose 
photometric analysis of high-$z$ quasar host galaxies suggests that at 1.7 
\lax\ $z$ \lax\ 4.5 the ratio of BH mass to bulge stellar mass is a factor of 
4--6 higher than the local value.  Of course, without 

\begin{figure*}[t]
\centerline{\psfig{file=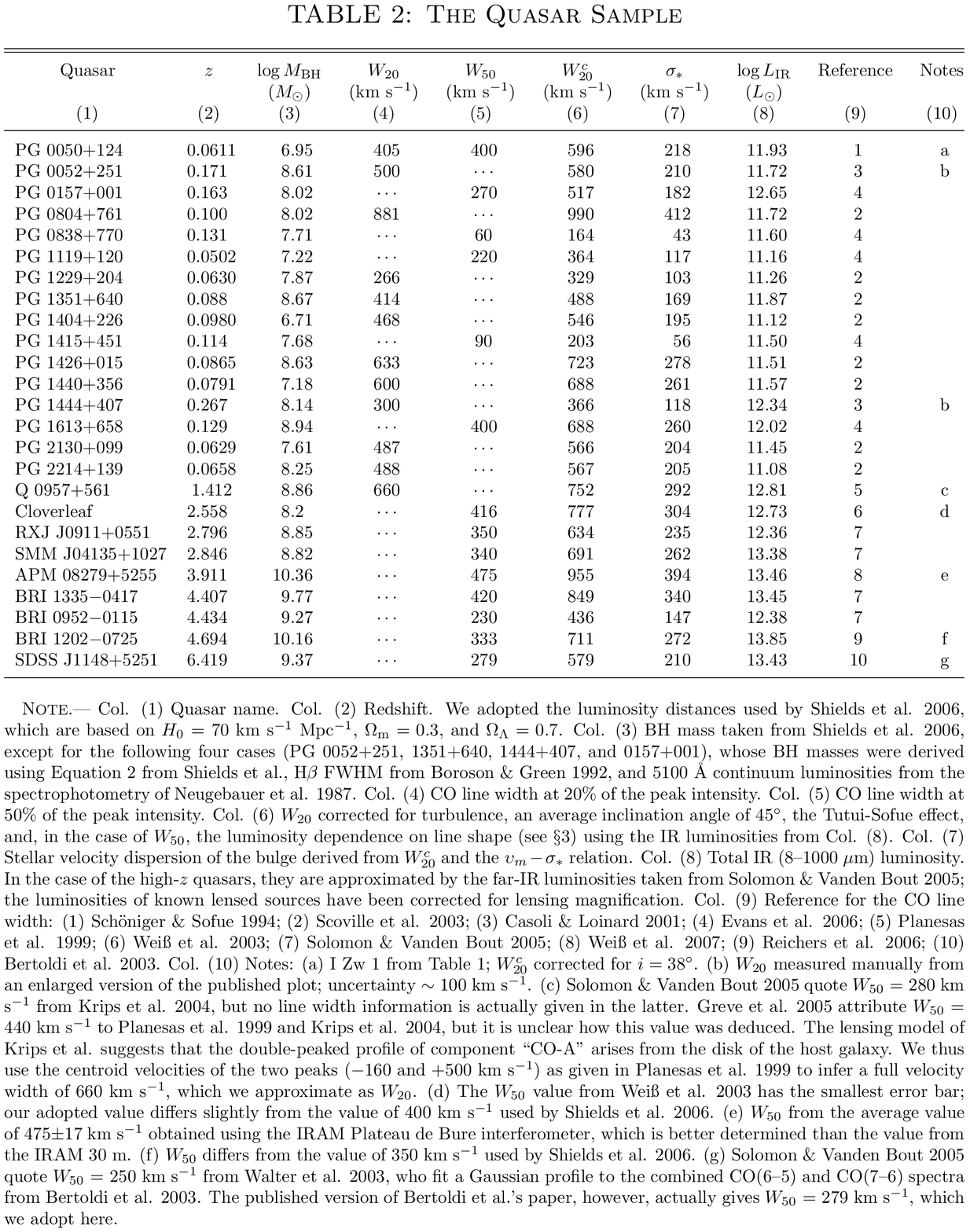,width=18.5cm,angle=0}}
\end{figure*}

\noindent 
actual observational 
constraints on inclination angles, it is impossible to disprove the prosaic 
alternative that these sources are simply biased toward more face-on 
orientations.  To shift the ensemble of high-$z$ points back to the local 
\msigma\ relation would require the average inclination angle to be 
$\langle i \rangle = 30$\deg\ instead of 45\deg, less extreme than 
$\langle i \rangle = 15$\deg\ found by Wu (2007).  

\vskip 0.3cm
\psfig{file=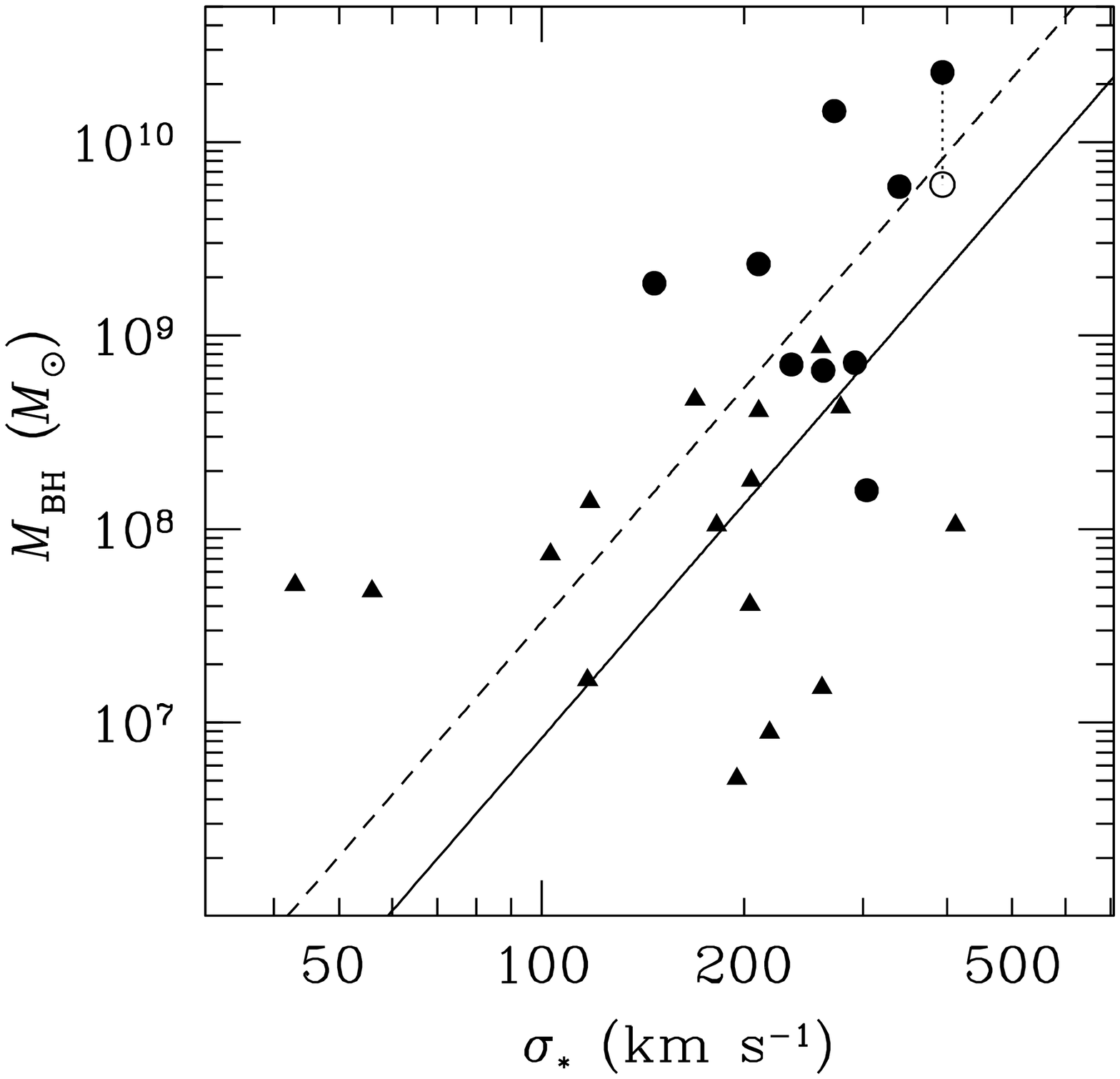,width=8.5cm,angle=0}
\figcaption[f3.eps]{
The \msigma\ relation for low-$z$ PG quasars ({\it triangles}) and
high-$z$ quasars ({\it circles}).  The BH masses were derived from the quasar
line width and luminosity using the virial technique, and velocity dispersion
comes from the CO line width.  The point marked with an open circle and
connected with a dotted line shows the new location of APM~08279+5255 with
its BH mass adjusted to the revised luminosity lensing magnification factor of
Wei\ss\ et al.  (2007).  The solid line shows the \msigma\ relation of
Tremaine et al. (2002) for local inactive galaxies, and the dashed line shows
an offset of 0.6 dex in mass from the local relation.
\label{f3}}
\vskip 0.3cm

\noindent
Despite this impasse, the 
good agreement between the present kinematic analysis and the photometric 
study of Peng et al. (2006b) lends credence to the proposition that at high 
redshifts the growth of the BH outpaced the growth of the bulge.

We end with a few cautionary remarks, which we hope can serve as a catalyst for 
future work.  First, we should recognize that BH masses derived from the 
virial method carry substantial uncertainty.  This is especially so for masses
that rely on the C~{\sc IV} \lamb 1549 line, which might be biased with 
respect to masses obtained from H\bet\ (Baskin \& Laor 2005).  It would be 
highly desirable to acquire high-quality near-IR spectra of all CO-detected 
high-$z$ quasars in order to reevaluate their BH masses using an internally
consistent, homogeneous set of line measurements based on either H\bet\ or 
Mg~{\sc II} \lamb 2800 (e.g., Barth et al. 2003).  A number of the high-$z$ 
quasars are known or suspected to be lensed.  More accurate lensing models 
would help to bolster our confidence in their intrinsic luminosities, which 
enter crucially into the BH mass estimate via the size-luminosity relation. 
Two of the objects in Table~2 have uncomfortably large BH masses ($M_\bullet > 
10^{10}$ \solmass), larger than any directly measured in the local Universe.  
BRI~1202$-$0725 is not known to be lensed, but APM~08279+5255 is.  Wei\ss\ 
et al. (2007) recently revised the luminosity magnification factor of 
APM~08279+5255 from 7, the value given in Solomon \& Vanden~Bout (2005) that 
Shields et al. (2006) adopted, to 100.  Since the virial mass scaling relation 
used by Shields et al. depends on $L^{0.5}$, reducing the luminosity by a 
factor of 14 lowers the BH mass by a factor $\sim 4$ (see Fig.~3).  At the 
moment we do not know how many of the other lensed objects may be similarly 
affected.  

The correction for line shape that we advocate in this study (Eq.~3) depends
on the total IR luminosity, but we have not specified what actually powers 
this luminosity.  Many IR-luminous galaxies are composite systems whose 
bolometric luminosity comes from a mixture of AGN heating and star formation.  
If the empirical line shape correction is driven largely by the starburst 
component to $L_{\rm IR}$,  then we may have conceivably overcorrected the 
CO line widths for the quasars, whose IR luminosity is at least partly, and 
some would argue predominantly (Ho 2005), due to AGN heating.  This issue 
can be addressed by obtaining quantitative estimates of the star formation 
rate in these systems, for instance using the technique proposed by Ho \& Keto 
(2007).  A more straightforward remedy, of course, is to simply take better 
CO spectra so that $W_{20}$ can be directly measured.

Finally, the apparently excellent agreement between our results and those of 
Peng et al. (2006b) is somewhat misleading.  High-redshift galaxies have 
steeper gravitational potential wells, and so a given \sigstar\ corresponds 
to a {\it smaller}\ bulge stellar mass (Robertson et al. 2006).  The factor of 
$\sim$4 increase in BH mass for high-$z$ quasars compared to the local 
\msigma\ relation, therefore, actually results in an ever larger offset 
had the comparison been done in terms of stellar mass, which is the approach 
taken by Peng et al. (2006b).  On the other hand, Peng et al.'s analysis 
assumed a rather conservative mass-to-light ratio, one designed to yield the 
largest bulge mass.  In this sense the factor of 4--6 increase in the 
BH-to-bulge mass ratio found by Peng et al. should be viewed as a lower limit 
(C. Y. Peng 2007, private communications), which brings their results in 
closer agreement with ours.

\vskip 1.3cm
\section{Discussion and Summary}

The increasing realization that BHs and their energy feedback play a 
fundamental role in the life-cycle of galaxies places great urgency in 
devising effective methods to diagnose high-$z$ quasars and their hosts.  
Many approaches can be taken to tackle this problem, but a particularly 
direct one is to attempt to detect evolution in the well-known local scaling 
relations between BH mass and host galaxy properties.  The host galaxies of 
high-$z$ quasars, however, pose a formidable challenge for studying in stellar
light.  Not only must one contend with the enormous glare of the luminous 
AGN, but one is further handicapped by surface brightness dimming, band-pass
shifting, and the uncertain star formation history of the galaxy.  While the 
photometric properties of the stars can be studied under optimal conditions 
(e.g., in lensed systems; Peng et al. 2006b), stellar kinematical information 
will be much more difficult to come by, even in the foreseeable future.
Within this backdrop, observations of the large-scale cold interstellar medium 
of the host can serve as a powerful tool to probe not only its gas content 
but its global kinematics (Ho et al. 2007).  From experience with galaxies in 
the nearby Universe, even a measurement as crude as an integrated spectrum can 
yield an important physical parameter, namely the rotation velocity of the 
disk.  The recent progress in the detection of CO emission in high-$z$ quasars 
(Solomon \& Vanden~Bout 2005) presents an opportunity to explore this problem.  

Since the \msigma\ relation serves as the best local reference benchmark, we 
must devise a method to relate CO line widths to \sigstar.  Shields et al. 
(2006) tried a number of proxies, but were unconvinced as to which was most 
reliable, in large part because their calibration samples were very small and 
because each of the empirical correlations they employed has significant 
intrinsic scatter.  In the end they simply assumed \sigstar\ $\approx$ 
$W_{50}$(CO)/2.35.  Wu (2007), by contrast, directly compared $W_{50}$(CO) 
with \sigstar\ measurements for 33 Seyfert galaxies to arrive at a conversion 
between the two.   

We have investigated the CO line width calibration problem from a very 
different perspective.  We began by showing that normal galaxies obey a tight 
CO Tully-Fisher relation in the $K_s$ band, and that we can recover the more 
familiar \hi-based relation provided that we use the line width measured at 
20\% of the peak ($W_{20}$) instead of that measured at 50\% of the peak 
($W_{50}$).  The reduction in the scatter is especially impressive for 
IR-luminous galaxies, whose frequently disturbed morphologies, as a result 
of tidal interactions and mergers, produce a wide array of CO line shapes, 
and hence large differences between $W_{20}$ and $W_{50}$.  Indeed, we show 
that the line shape, as gauged by $W_{20}$/$W_{50}$, varies substantially 
and systematically with IR luminosity.  We further discuss additional 
corrections that need to be applied to the line widths in order to achieve 
the best possible results.  Having established $W_{20}$ to be superior to 
$W_{50}$ as a measure of maximum rotation velocity (\vm) of the disk, as well 
as an empirical guideline to convert $W_{50}$ to $W_{20}$ if the latter is 
not available, we then turn to the issue of how to relate $W_{20}$ (or 
equivalently \vm) to \sigstar.  Since \sigstar\ measures the bulge and \vm\ 
reflects the circular velocity of the halo, it is not immediately apparent how
the two should be related.  This issue was recently addressed by Ho (2007), 
who showed that although \vm\ does not correlated as tightly with \sigstar\ as 
had been thought previously, nevertheless galaxies over a wide range of 
Hubble types do follow a \vm--\sigstar\ relation.  This last step then 
completes the recipe to transform CO line widths to \sigstar.

With these tools in hand, we revisited the CO-based \msigma\ relation for 
quasars.  Low-redshift ($z$ \lax\ 0.2) quasars fall roughly on the \msigma\ 
relation of inactive galaxies, albeit with considerable scatter, but 
high-redshift ($z = 1.41-6.42$) quasars statistically lie displaced from the 
local relation.  Our finding is in qualitative agreement with the results of 
Shields et al. (2006), but the magnitude of the offset has been reduced with 
our more refined \sigstar\ estimates.  At a given \sigstar, high-$z$ quasars 
have BH masses larger by a factor of $\sim$4 relative to local galaxies.  
Although we cannot rigorously rule out the possibility that we are being 
misled by a population of quasars with biased orientations, we are encouraged 
by the fact that our result quantitatively agrees well with that based on the 
photometric analysis presented by Peng et al. (2006b).  The collective evidence
suggests that early in the life-cycle of galaxies the development of the bulge 
lags behind the growth of the central BH.  

\acknowledgements
This work was supported by the Carnegie Institution of Washington and by NASA 
grants from the Space Telescope Science Institute (operated by AURA, Inc., 
under NASA contract NAS5-26555).  Extensive use was made of the NASA/IPAC 
Extragalactic Database (NED), which is operated by the Jet Propulsion 
Laboratory, California Institute of Technology, under contract with NASA.  I 
thank Jenny Greene, Chien Peng, and Greg Shields for discussions and comments 
on this paper.  An anonymous referee provided very helpful suggestions.

 

\end{document}